\documentclass[fleqn,usenatbib]{mnras}

\usepackage{newtxtext,newtxmath}

\usepackage[T1]{fontenc}

\DeclareRobustCommand{\VAN}[3]{#2}
\let\VANthebibliography\thebibliography
\def\thebibliography{\DeclareRobustCommand{\VAN}[3]{##3}\VANthebibliography}

\usepackage{graphicx}	
\usepackage{amsmath}	
\usepackage{hyperref}

\usepackage[dvipsnames]{xcolor}

\usepackage{soul}

\title[\ion{H}{i} in M51 with FAST]{High sensitivity \ion{H}{i} image of diffuse gas and new tidal features in M51 observed by FAST}

\author[H. Yu et al.]{
Haiyang Yu,$^{1,2,4}$\thanks{E-mail: hyyu@nao.cas.cn}
Ming Zhu,$^{1,2,3,4}$\thanks{E-mail: mz@nao.cas.cn}
Jin-Long Xu,$^{1,3,4}$ 
Mei Ai,$^{1,3,4}$ 
Peng Jiang$^{1,3,4}$
and Yanbin Yang$^{5}$
\\
$^{1}$National Astronomical Observatories, Chinese Academy of Sciences, A20 Datun Road, Beijing 100101, China\\
$^{2}$University of Chinese Academy of Sciences, Beijing 100049, China\\
$^{3}$Guizhou Radio Astronomical Observatory, Guizhou University, Guiyang 550000, People's Republic of China\\
$^{4}$CAS Key Laboratory of FAST, National FAST, National Astronomical Observatories, Chinese Academy of Sciences, Beijing 100101, China\\
$^{5}$GEPI, Observatoire de Paris, CNRS, Place Jules Janssen 92195, Meudon, France.
}

\date{Accepted XXX. Received YYY; in original form ZZZ}

\pubyear{2023}

\begin{document}
\label{firstpage}
\pagerange{\pageref{firstpage}--\pageref{lastpage}}
\maketitle

\begin{abstract}
We observed the classical interacting galaxy M51 with FAST and obtain high sensitivity \ion{H}{i} image with column density down to 3.8 $\times$ 10$^{18}$ cm$^{-2}$. 
In the image we can see a diffuse extended envelope around the system and several new tidal features. 
We also get a deeper look at M51b's probable gas, which has an approximated velocity range of 560 to 740 km s$^{-1}$ and a flux of 7.5 Jy km s$^{-1}$.
Compared to the VLA image, we observe more complete structures of the Southeast Tail, Northeast Cloud and Northwest Plume, as well as new features of the Northwest Cloud and Southwest Plume.
M51's most prominent tidal feature, the Southeast Tail, looks very long and  broad, in addition with two small detached clouds at the periphery.
Due to the presence of optical and simulated counterparts, the Northwest cloud appears to be the tail of M51a, while the Northwest Plume is more likely a tidal tail of M51b.
The large mass of the Northwest Plume suggests that M51b may have been as gas-rich as M51a before the interaction.
In addition, the formation process of the Northeast Cloud and Southwest Plume is obscured by the lack of optical and simulated counterparts.
These novel tidal features, together with M51b's probable gas, will inspire future simulations and provide a deeper understanding of the evolution of this interacting system.
\end{abstract}




\section{Introduction}

Galaxy interactions and mergers caused by gravity are believed to play an important role in galaxy evolution, with more and more detailed features have been observed from nearby galaxies, such as warping of galactic discs, bridges and tails between galaxies, multiple massive black holes in galactic centre, and so on.
Understanding of interaction histories and futures from galaxies' features and physical processes is critical. 
M51 is a one of well known interacting systems in the local Universe, consisting of a nearly face-on spiral galaxy M51a (NGC5194), and its companion M51b (NGC5195).
Since the grand-design spiral arm of M51a was discovered by Lord Rosse in 1845, many works have focused on this interacting galaxy pair in both observation and theory to study galaxy evolution, tidal tails and star formation \citep[e.g.][]{Toomre1981, Bastian2005, Calzetti2005, Kennicutt2007}.

M51 multi-wavelength images from X-ray to radio have been observed to study its morphology and kinematics \citep[e.g.][]{Rots1990, Dale2005, Watkins2015, Heyer2022}.
M51a is a Sbc galaxy with a weak Seyfert 2 active galactic nucleus (AGN) in its centre, while M51b is an early-type SB0 galaxy \citep{deVaucouleurs1995}.
A deep X-ray image using Chandra shows the distribution of X-ray point sources.
The luminosity function suggests that M51a is dominated by high-mass X-ray binaries, while M51b is dominated by low-mass X-ray binaries \citep{Kuntz2016}.
In optical bands, the magnificent spiral arms of M51a, three-pronged structure of M51b and bridge between galaxies are system's iconic structures.
In deeper observation, some extended faint features can be seen, such as the Northwest, Northeast and South Plume \citep{Watkins2015}.
\citet{Watkins2018} discovered a diffuse ionized gas cloud with no stellar counterpart north of M51.
    They believe that the cloud was ejected from the centre of M51 due to its high metallicity, and then ionized by a hard ionization source.
For the observations of the \ion{H}{i} 21 cm emission line, \citet{Weliachew1973} observed a position angle of ~22$^\circ$ for the morphological major axis of NGC5194 and ~-8$^\circ$ for the dynamically major axis with the two element interferometer of the Owens Valley Radio Observatory. 
They also found a large non-circular velocity in the northeast corner due to the interaction with NGC5195, and a potential southeast tidal tail structure.
The high-resolution \ion{H}{i} observation with the Very Large Array (VLA) show that the inner \ion{H}{i} is the spiral structure overlapping with the star-forming regions, and the outer part is a counter-rotating tail extending from the south to the northeast that does not look like an extension of the optical tail, but rather a independent structure \citep{Rots1990}.

As more and more studies on the structures and dynamics of M51 have been conducted, the simulation work is also developing in parallel \citep[e.g.][]{Toomre1972, Appleton1986, Burkhead1978, Durrell2003, DeLooze2014}. 
As a start, \citet{Toomre1972} used a simple numerical model to construct bridge and tail of M51 in optical images. 
Then \citet{Hernquist1990} and \citet{Howard1990} made a self-gravity numerical simulation to obtain not only the outer regions structure similar to the non-self-gravity model, but also the central spiral structure.
\citet{Howard1990} also simulated the extended \ion{H}{i} tail by using a bound orbit.
These early models mostly assumed that there was only one encounter between this galaxy pair.
To reveal more precise structures and dynamics, some models suggested a multiple-encounter scenario between two galaxies based on N-body and hydrodynamic simulations \citep{Salo2001, Theis2003, Dobbs2010}.

Although M51 has been studied for decades as an excellent interacting system, there are still many questions that need to be explored.
The deep optical images show faint distributions of stars to the north and northwest of M51 without gas, and the \ion{H}{i} southeast tail has no stellar counterpart.
For the history of this system, the number of galaxy encounters and whether there are more merger events still need to be studied by simulation.
Deeper \ion{H}{i} observations would be beneficial, and the Five-hundred-meter Aperture Spherical radio Telescope's (FAST) great sensitivity makes this possible. 
FAST observations have revealed several novel structures, such as the filament of M106 \citep{Zhu2021}, the extra-planar clouds and tail of M101 \citep{Xu2021}, and a 0.6 Mpc structure of the famous compact galaxy group "Stephen's Quintet" \citep{Xu2022}. 
A vast range of \ion{H}{i} features in interacting galaxy systems, such as tails, plumes, clouds and filaments \citep{Koribalski2020}, have been discovered, providing numerous references for the new study of M51.
In this paper, a high sensitivity \ion{H}{i} image observed by FAST is presented.
Newly discovered gas features and dynamics can be used to constrain models for studying galaxy interactions and evolution.
In section ~\ref{sec:obs}, we introduce observation and data processing.
In section ~\ref{sec:results}, we present \ion{H}{i} image and its discoveries.
Finally, in section ~\ref{sec:discussion}, we explore the possible origin of the discoveries.

\section{Observations and data processing}
\label{sec:obs}

\begin{figure}
	\includegraphics[width=\columnwidth]{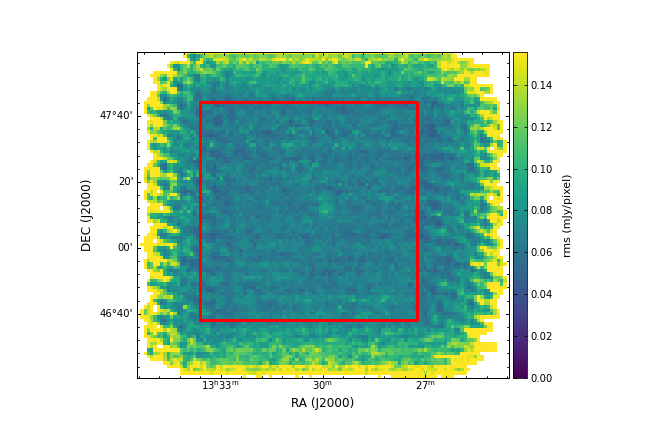}
    \caption{The RMS distribution map covering the whole observation area. 
    The red box marks the area used for subsequent analysis of M51.}
    \label{fig:rms}
\end{figure}

We observed M51 twice with the FAST on December 2, 2021. 
FAST \citep{Jiang2019, Jiang2020, Qian2020} is a single-dish radio telescope located in Guizhou, China, and its aperture is 500 m, while the illuminated aperture is 300 m under the coordinated control of an active reflector system and feed support system.
The 19-beam receiver \citep{Dunning2017} is used which covers a frequency range from 1050 MHz to 1450 MHz and works in four polarization modes.
For \ion{H}{i} observations, two linear polarization data are used in subsequent data processing.
For the FAST digital spectral-line backend, we select the Spec(W) spectrometer that covers 500 MHz with 65536 channels for wide band. 
As a result, the frequency and velocity resolution are 7.6 kHz and 1.6 km s$^{-1}$ at 1.4 GHz.
The half-power beamwidth (HPBW) of 19-beam receiver each beam at 1.4 GHz is approximately 2.9' and the system temperature is around 25 K, which is related to the zenith angle (ZA).
The pointing errors are less than 16" with FAST at full gain (ZA $\leq$  26.4$^\circ$) and its root mean square (RMS) is 7.9".
The observation mode we choose is MultiBeam on-the-fly (OTF) mapping with a 23.4$^\circ$ rotation of the 19-beam receiver platform, 10.3' scanning separation and 15"s$^{-1}$ scanning speed.
In this case, the scan along declination direction can satisfy Nyquist sampling and maximize sky coverage.
For the region of M51, a total of 4470 s is required for one OTF map, and we map this region twice to increase the sensitivity.

For data processing, we mainly adopted the \ion{H}{i}FAST pipeline developed by Jing et al. (2023, in preparation) for \ion{H}{i} spectral-line observed by FAST. 
Here we briefly describe the processing procedures.
First, we converted the instrument unit to the antenna temperature by injecting a high-intensity noise diode (~12 K) for 1 s every 32 s and using the following equation,

\begin{equation}
\begin{aligned}
    &T^{\mathrm{on}}_\mathrm{A}=\frac{P_{\mathrm{on}}}{P_{\mathrm{on}}-P_{\mathrm{off}}}T_{\mathrm{cal}}-T_{\mathrm{cal}}\\
    &T^{\mathrm{off}}_\mathrm{A}=\frac{P_{\mathrm{off}}}{P_{\mathrm{on}}-P_{\mathrm{off}}}T_{\mathrm{cal}}
    \label{eq:tonoff}
\end{aligned}
\end{equation}
where $T^{\mathrm{on}}_\mathrm{A}$ and 
$T^{\mathrm{off}}_\mathrm{A}$ are antenna temperatures while noise diode is on and off.
$P_{\mathrm{on}}$ and $P_{\mathrm{off}}$ are instrument units.
$T_{\mathrm{cal}}$ is noise diode temperature measured before.
Then the feed system coordinates are used to calculate the right ascension (RA) and declination (DEC) coordinates for each sampling point.
The next step is to convert antenna temperature to flux density via telescope gain, which is about 16 K Jy$^{-1}$ and is related to the ZA and beam.
The asymmetrically reweighted penalized least-squares (arPLS) algorithm is used for baseline correction, which can flatten the baseline well and hardly affect the signal.
Before creating the data cube, the spectral lines need to undergo velocity correction, frame conversion (horizontal to heliocentric) and polarization merging.
Finally, the calibrated spectral lines are gridded into a image with the spatial resolution of 1' and a standard FITS format data cube is created.
In order to obtain higher quality image, we performed standing wave removal, Hanning smoothing and baseline correction to the spectral lines of the data cube.
After the Hanning smooth, the velocity resolution of the spectra is 4.8 km s$^{-1}$.
Fig.~\ref{fig:rms} depicts the RMS distribution of the final data.
Because of the small number of sampling points, the RMS at the edge area is higher. 
The RMS of the spectral lines per pixel in the central region is approximately 0.07 mJy. 
We observed a sky area of around 1.8$^\circ$ $\times$ 1.6$^\circ$. For the following study, we cut the inner region of 1.1$^\circ$ $\times$ 1.1$^\circ$ (the red boxed area).

\section{Results}
\label{sec:results}

\begin{figure*}
	\includegraphics[width=2\columnwidth]{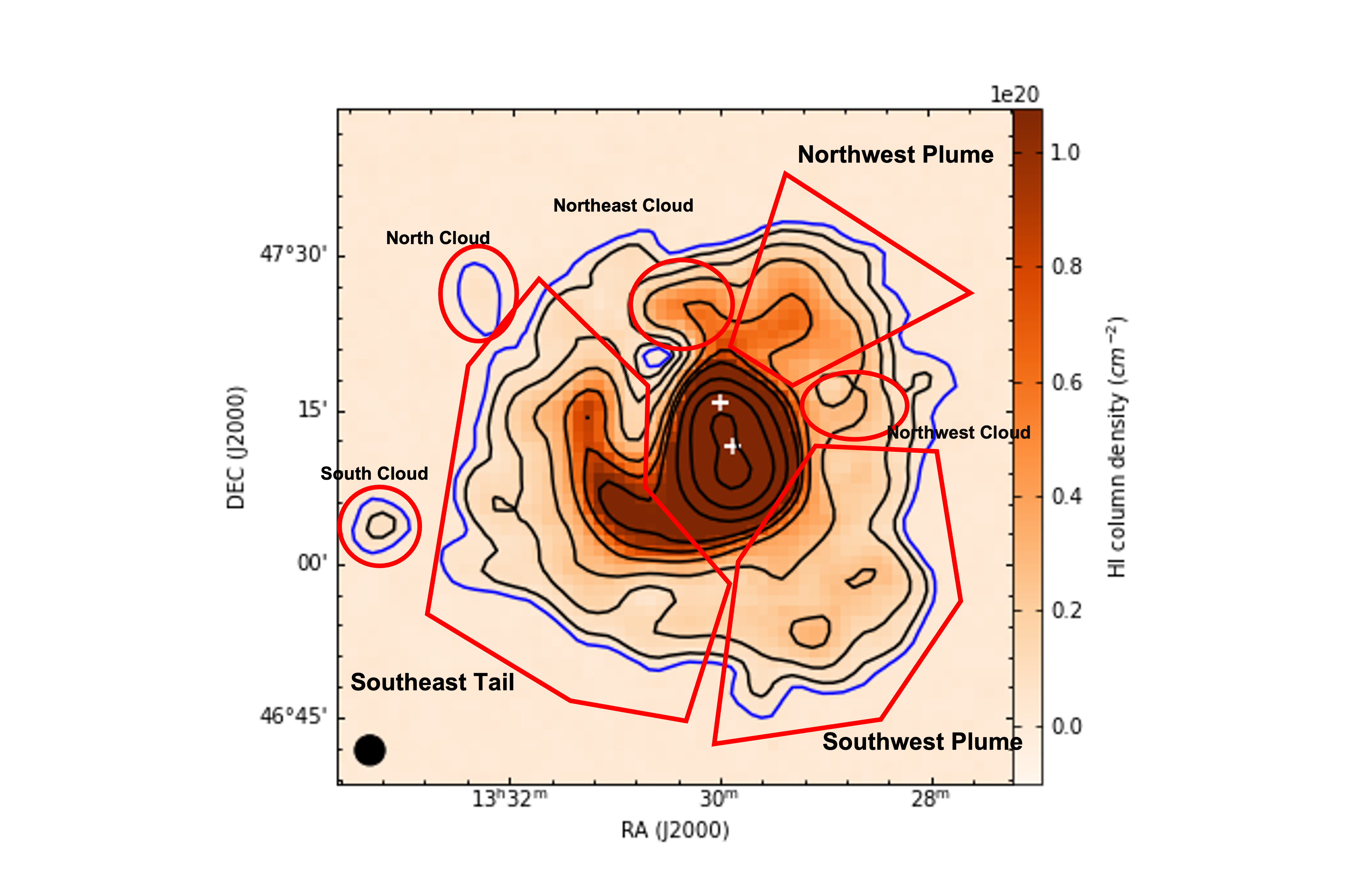}
    \caption{The \ion{H}{i} column density distribution of M51 with contours integrated over 280-750 km s$^{-1}$.
    The contour levels are 3.8 (5$\sigma$, blue line), 6.4, 12.8, 25.6, 51.2, 76.8, 102.4, 204.8, 409.6, and 819.2 $\times$ 10$^{18}$ cm$^{-2}$.
    Two white crosses mark the locations of M51a (bottom) and M51b (top), respectively.
    The red lines marks the \ion{H}{i} features as listed in detailed in Table~\ref{tab:feature_table}.
    The HPBW of FAST is shown in the bottom-left corner.}
    \label{fig:moment0}
\end{figure*}

\begin{figure}
	\includegraphics[width=\columnwidth]{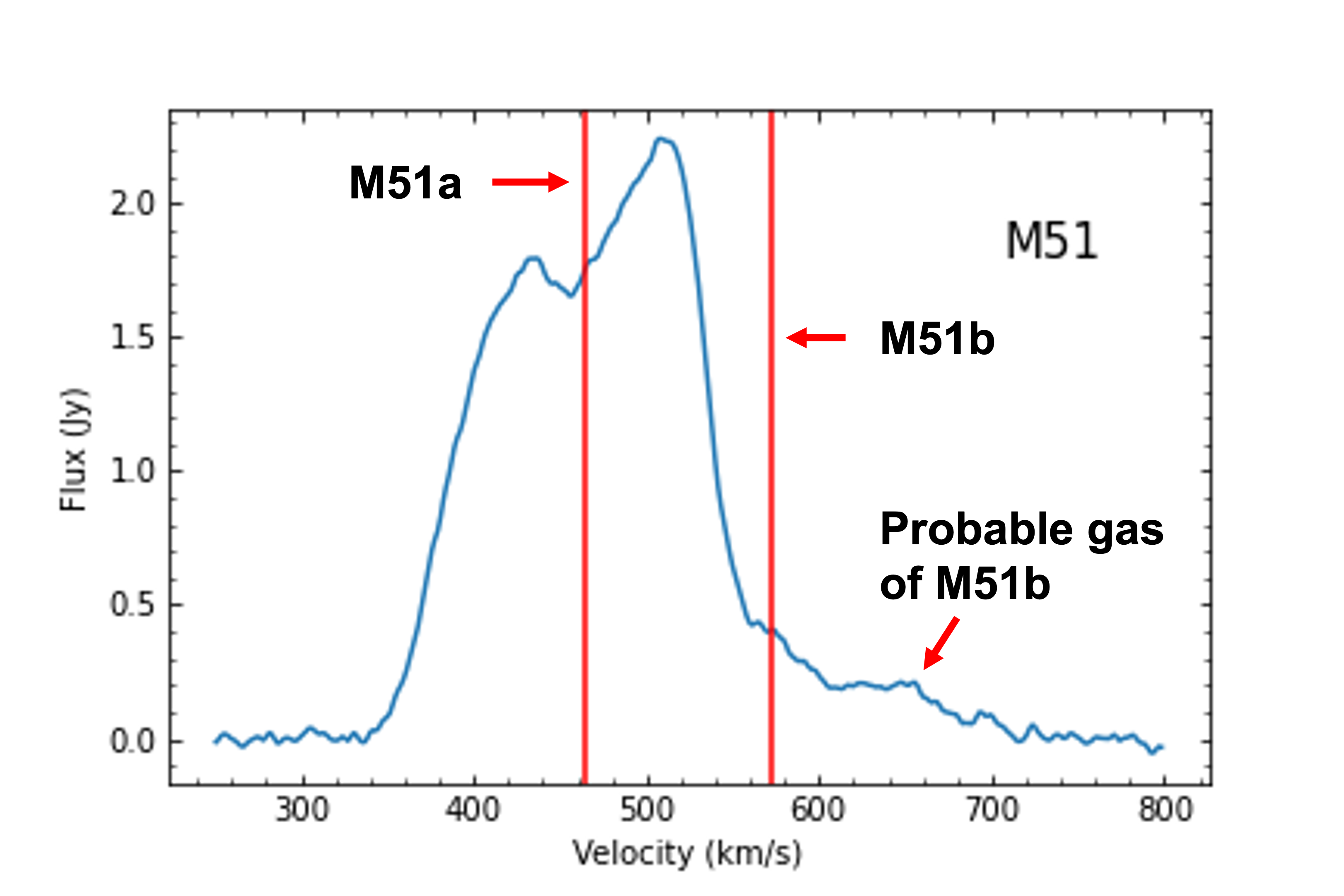}
    \caption{Global \ion{H}{i} profile for M51.
    The two red vertical lines mark the velocity of M51a and M51b respectively.}
    \label{fig:spec}
\end{figure}

\begin{figure*}
	\includegraphics[width=2\columnwidth]{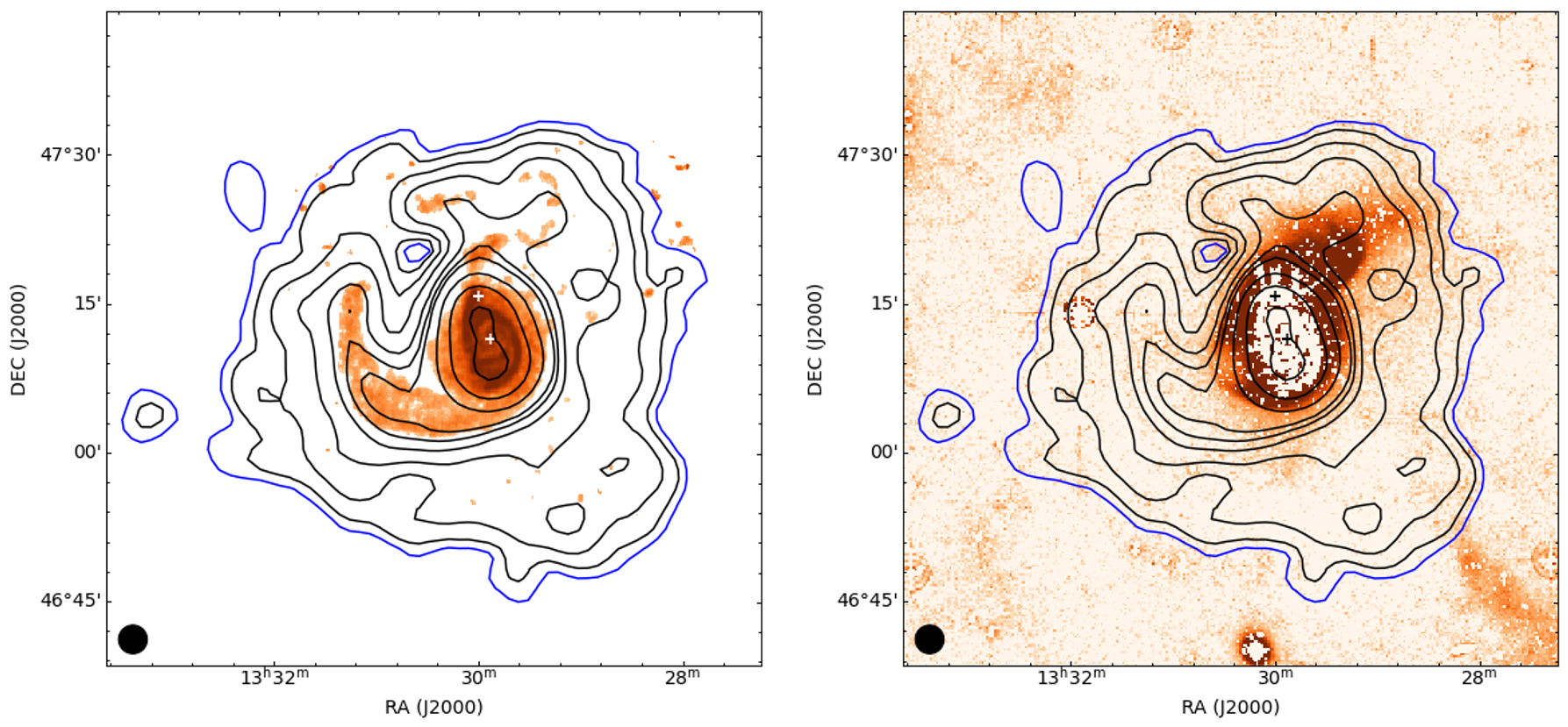}
    \caption{The \ion{H}{i} column density contours which levels are consistent with that in Fig.~\ref{fig:moment0} overlaid on the VLA \ion{H}{i} image \citep[left,][]{Rots1990} and deep B band image \citep[right,][]{Watkins2015}.
    To better show the faint features, the deep optical image has been median binned and masked the high brightness areas.
    Two crosses mark the locations of M51a (bottom) and M51b (top), respectively.
    FAST's HPBW is indicated in the bottom-left corner.
    }
    \label{fig:VLA_opt}
\end{figure*}

Fig.~\ref{fig:moment0} shows the \ion{H}{i} column density distribution of M51 integrated over 280-750 km s$^{-1}$.
The minimum detectable column density is 3.8 $\times$ 10$^{18}$ cm$^{-2}$ (5$\sigma$).
Due to FAST's high sensitivity and M51's face-on morphology, we can identify many different structures on the outer side of M51's high density gas disc, as marked by the red lines.
A long Southeast tail extends from the south to the northeast, with two faint gas clouds South and North cloud further east.
The Southwest and Northwest Plume are two diffuse extended gas on the west side.
There are also two clouds, the Northeast and Northwest Cloud, on either side of the Northwest Plume.
The flux from FAST in the area overlapping with the VLA observation is approximately 225 Jy km s$^{-1}$, which is matched with the previous data \citep[210 Jy km s$^{-1}$ observed by VLA, and 227 Jy km s$^{-1}$ observed by single-antenna;][]{Rots1990,Rots1980}.
The total flux of M51 region calculated from Fig.~\ref{fig:moment0} is 333 Jy km s$^{-1}$.
The extra flux mainly comes from the gas at the outskirts of M51 not covered by the VLA map.
Fig.~\ref{fig:spec} shows the global \ion{H}{i} profile for M51 integrated over the whole region of Fig.~\ref{fig:moment0}. 
The profile is basically consistent with the previous single-antenna observation data \citep{Rots1980}.
There are two peaks near the M51a's velocity and a tail near M51b's, which we will discuss in detail below as a probable gas component for M51b.
VLA observed the Southeast tail of M51 as well as a part of the Northwest Plume and Northeast Cloud, as seen in the left panel of Fig.~\ref{fig:VLA_opt}. 
However, the structure of the system's northwestern part is so fragmented that it is difficult to analyse its origin specifically. Fortunately, such problem is fixed by the FAST image. 
In the FAST map, we find not only the more complete Southeast Tail, Northwest Plume, and Northeast Cloud, but also some other new features such as the South Cloud, North Cloud, Southwest Plume, and Northwest Cloud.
In the right panel of Fig.~\ref{fig:VLA_opt}, the overlay of the \ion{H}{i} intensity contours on the deep optical image shows the overlap of the Northwest Plume and Northwest Cloud with the faint stellar features, as well as the offset of the Southeast Tail, South Cloud, North Cloud, Southwest Plume, and Northeast Cloud.
Fig.~\ref{fig:moment1} is the moment-1 map of M51, which shows the irregularity and complexity of the velocity field of M51 as a result of various effects, making it difficult to analyse.
The position-velocity (PV) diagram (Fig.~\ref{fig:pv}) and channel map (Fig.~\ref{fig:channel_map}) are good tools for studying gas properties and dynamics.
In the PV diagram, we extract six slices of the cube along different directions with a width of two pixels for analysis and mark the corresponding gas features and probable gas of M51b.
The channel map's velocity range is 280-750 km s$^{-1}$ and the integration range for each panel is 13.05 km s$^{-1}$. 
Next, we will discuss the \ion{H}{i} features marked in Fig.~\ref{fig:moment0} in details in conjunction with the velocity map, PV diagram and channel map.
The properties of \ion{H}{i} features can be seen in Table~\ref{tab:feature_table}.

\begin{table*}
	\centering
	\caption{properties of \ion{H}{i} features}
	\label{tab:feature_table}
	\begin{tabular}{lccc} 
		\hline
		Region & Flux (Jy km s$^{-1}$) & Mass(10$^{6}$ M$_{\odot}$) & Velocity (km s$^{-1}$)\\
		\hline
		Southeast Tail & 60 & 1042.4 & 320-540\\
		North Cloud & 0.62 & 10.8 & 440-510\\
		South Cloud & 0.43 & 7.5 & 380-430\\
        Southwest Plume & 21 & 364.8 & 370-570\\ 
        Northwest Plume & 24.8 & 430.9 & 510-710\\
        Northwest Cloud & 3.5 & 60.8 & 460-560\\
        Northeast Cloud & 6 & 104.2 & 480-600\\
        Probable M51b's gas & 7.5 & 130.3 & 560-740\\
		\hline
	\end{tabular}
\end{table*}

\subsection{\ion{H}{i} features outside the M51 disc}

The high sensitivity map of FAST reveals more diffuse and faint gas surrounding the Southeast Tail of M51. 
The FAST observed tail is wider and longer than that revealed by the VLA.
It is up to 40' in projected length \citep[or 100 kpc with a distance of 8.58 Mpc;][]{McQuinn2016}, and its width is greater than 15' (or 37 kpc).
The flux of the tail is about 60 Jy km s$^{-1}$, which is 18\% of the whole M51 flux.
The mass is about 1.04 $\times$ 10$^{9}$ M$_{\odot}$, which is twice of that reported by VLA, by using the following equation,

\begin{equation}
    M_{\ion{H}{i}}=2.36 \times 10^5 d^2\int S_\nu \mathrm{d} v
    \label{eq:mass}
\end{equation}
where the mass unit is solar mass.
$d$ is the distance in Mpc.
$S_\nu$ is the flux at different frequencies, and its unit is Jy.
The velocity is in km s$^{-1}$.

\begin{figure*}
	\includegraphics[width=2\columnwidth]{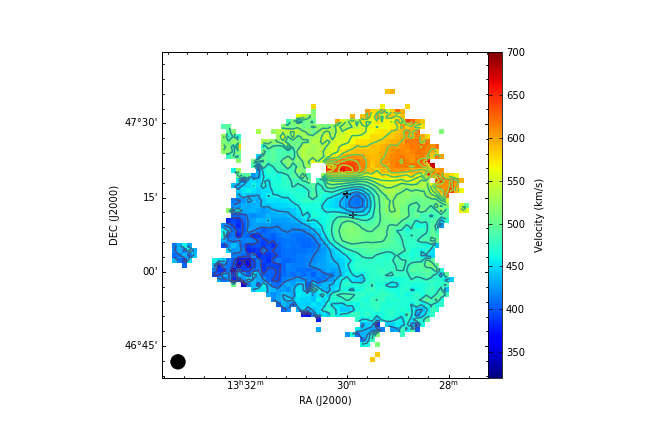}
    \caption{Moment-1 map (velocity field) of M51 with contours and colourscale.
    The contour levels go from 340 to 680 km s$^{-1}$ in steps of 20 km s$^{-1}$.
    M51a (bottom) and M51b (top) are marked by two black crosses, respectively.
    The HPBW of FAST is indicated in the bottom-left corner.}
    \label{fig:moment1}
\end{figure*}

\begin{figure*}
	\includegraphics[width=2\columnwidth]{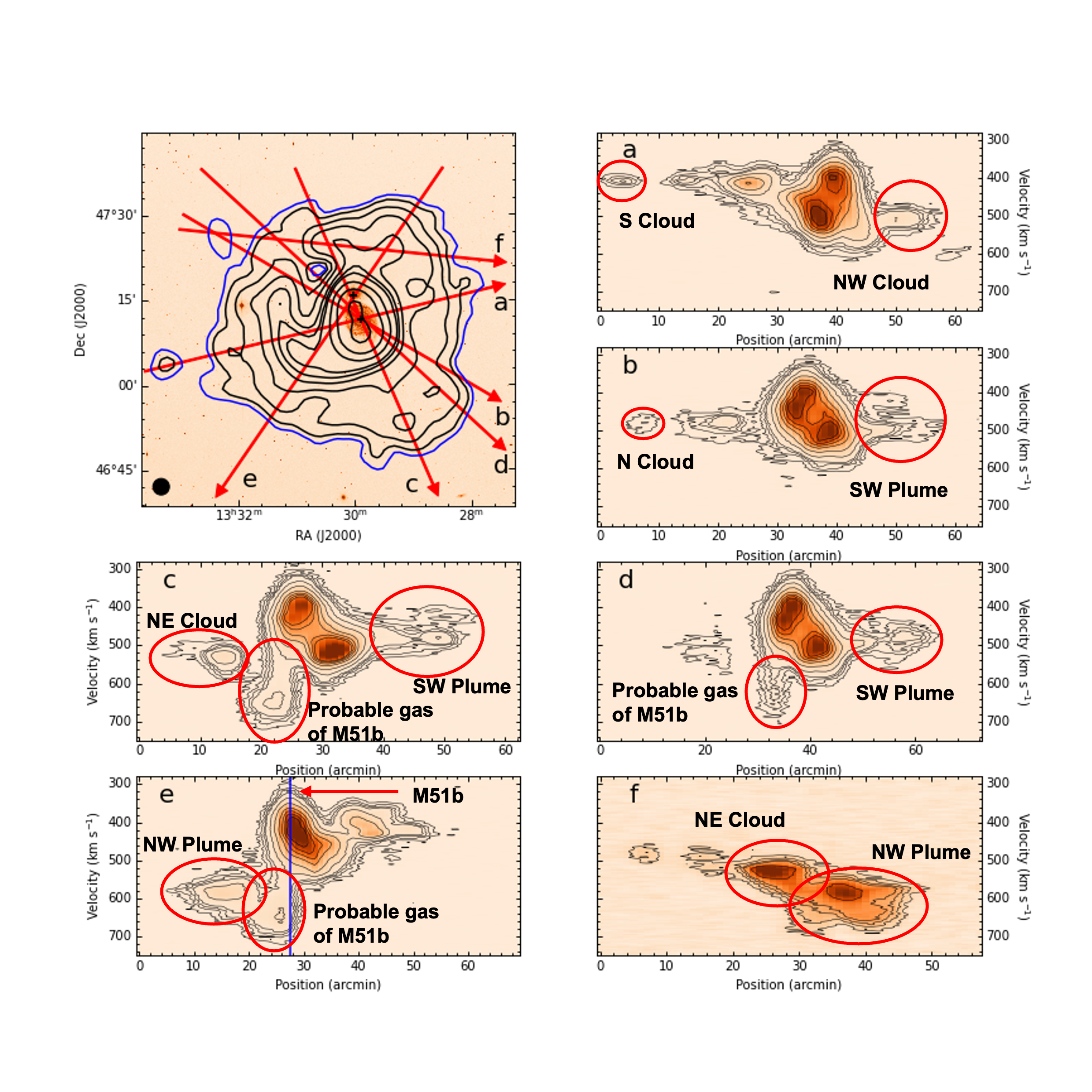}
    \caption{The left-top panel shows the DSS B-band optical image with FAST \ion{H}{i} column density contours. 
    The two black crosses are the centres of the optical images of M51a (bottom) and M51b (top), respectively. 
    The six red arrows mark the direction of the a-f positon-velocity diagrams. 
    The contours in PV diagrams begin at 0.21 mJy (3$\sigma$) and go on to 0.42, 0.63, 0.84, 1.68, 3.36, 6.72, 10.08, 13.44, and 16.8 mJy.
    The tidal features and M51b's probable gas are marked in subplots a-f, and the blue vertical line in e marks the centre of M51b.
    }
    \label{fig:pv}
\end{figure*}

There are two isolated clouds to the east of the tail.
The mass of North Cloud is about 10.8 $\times$ 10$^{6}$ M$_{\odot}$, and South Cloud's is 7.5 $\times$ 10$^{6}$ M$_{\odot}$.
In Fig.~\ref{fig:moment1} and Fig.~\ref{fig:pv} (a), (b), we can see that these two clouds' velocities are nearly identical to the velocities of their approaching tails' region.
The consistency in position and velocity suggest that these two clouds are fragments from the tail.
Their distances to the centre of M51a galactic disc are 30' or 75 kpc for the North Cloud and 35' or 87 kpc for the South Cloud, respectively.

\begin{figure*}
	\includegraphics[width=2\columnwidth]{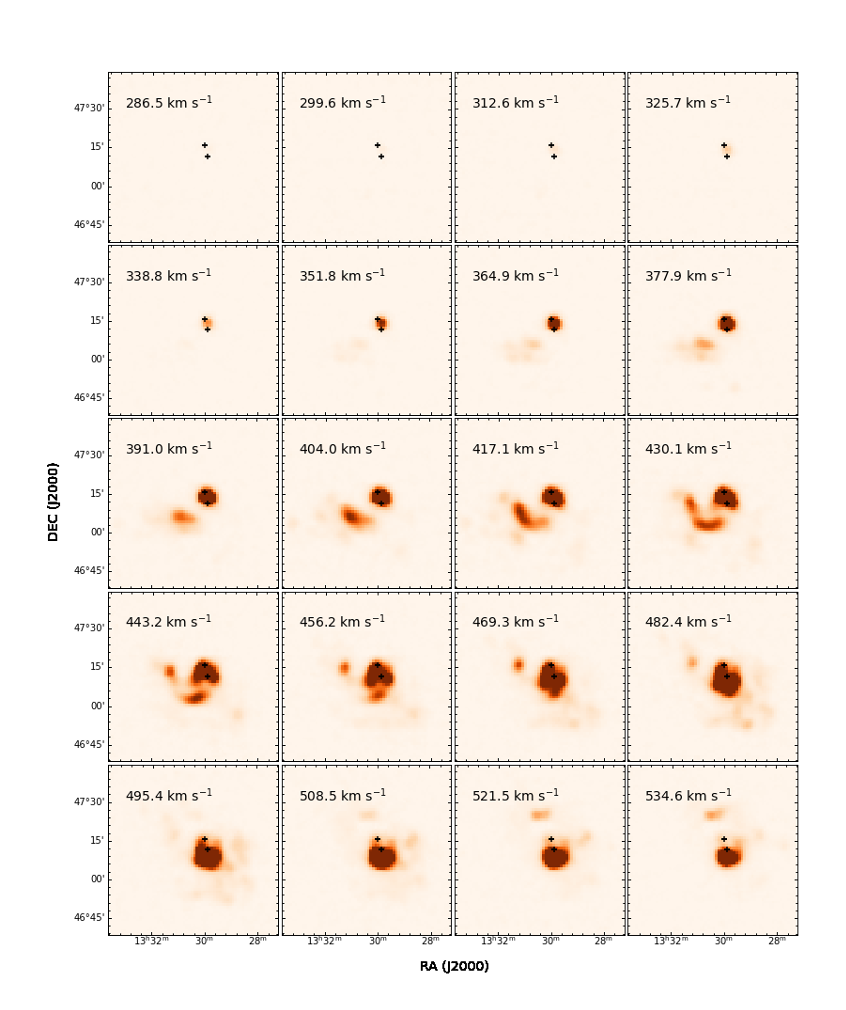}
    \caption{Channel map of M51. 
    The total velocity range is 280-750 km s$^{-1}$. 
    The integration range for each panel is 13.05 km s$^{-1}$ and the central velocity of the panel is indicated in the left-top corner. 
    The two black crosses are the centres of M51a (bottom) and M51b (top), respectively.}
    \label{fig:channel_map}
\end{figure*}

\begin{figure*}
    \setcounter{figure}{6}
	\includegraphics[width=2\columnwidth]{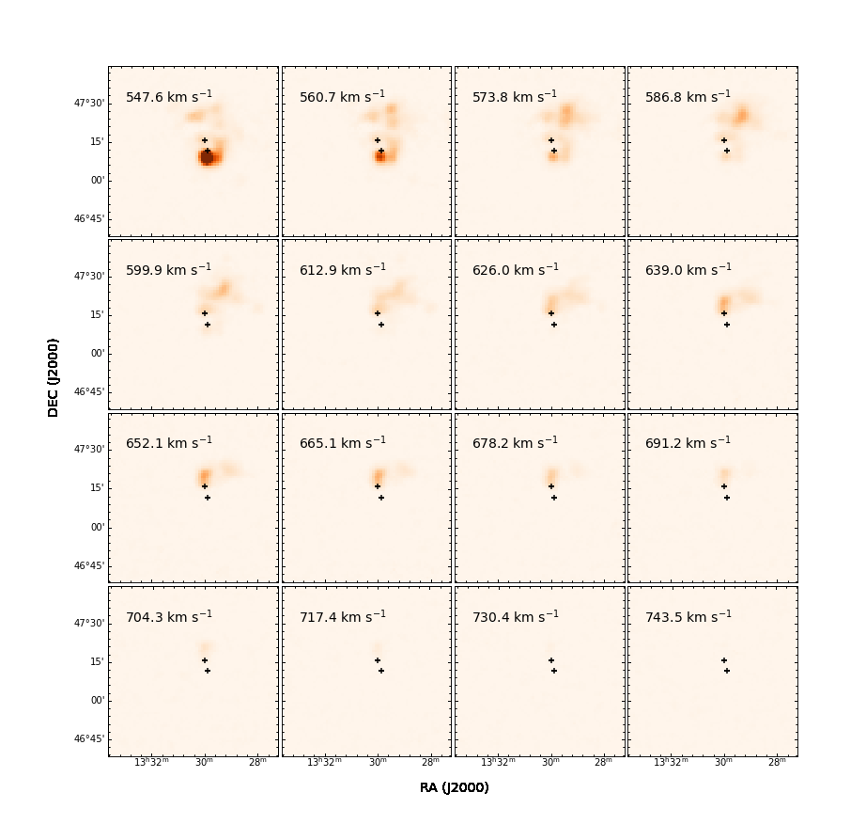}
    \caption{(continued)}
    \label{fig:channel_map}
\end{figure*}

The Southwest plume is an extended feature.
It is very diffuse and wraps the entire southwest side of M51a.
In Fig.~\ref{fig:pv} (b), (c) and (d), we can see that the plume's velocity is an extension of the M51a disc's southwest edge velocity.
What makes this plume special is the presence of two velocity features.
The bimodal structure can be seen in both Fig.~\ref{fig:pv} (b) and (c).
This plume can first be seen in Fig.~\ref{fig:channel_map} at centre velocity of 377.9 km s$^{-1}$ on the southern outskirts of M51a.
As the velocity rises, a stream of gas is present to the south of M51a. 
The gas associated to M51a disc's southwest and western clouds are then visible when the velocity approaches 500 km s$^{-1}$.
The total flux of Southwest Plume is about 21 Jy km s$^{-1}$.
The complexity of the velocity and position distribution of this plume probably stems from the multiple-encounter of M51 system.

The Northwest Plume is another newly observed extended feature.
In Fig.~\ref{fig:pv} (e), we can see that there appear to be two components of the high velocity gas in the northwestern of M51, one inside and one outside the disc of M51a, and here we define the part outside the disc as the Northwest Plume.
In Fig.~\ref{fig:moment1}, it can be seen that the Northwest Plume is the fastest component of the M51 system and has a large velocity span, from 510 km s$^{-1}$ to 710 km s$^{-1}$. 
Fig.~\ref{fig:channel_map} shows that as the velocity increases, the gas of plume extends from the north outskirts to the northwest of M51.
In this region, optical counterparts dominated by old stars and a high metallic ionized gas cloud (velocity is 637 $\pm$ 13 km s$^{-1}$) have also been discovered \citep{Watkins2015,Watkins2018}.

The Northwest Cloud is the counterpart of a slight protrusion in deep optical image \citep{Watkins2015}, sandwiched between two plumes.
It has a flux of around 3.5 Jy km s$^{-1}$, which equals a mass of 6.08 $\times$ 10$^{7}$ M$_{\odot}$.
In Fig.~\ref{fig:pv} (a), the Northwest Cloud has a velocity range of 460-560 km s$^{-1}$, consistent with the northwest side of M51a disc.

The Northeast cloud is a high density cloud, part of which was detected by VLA.
It has a flux of about 6 Jy km s$^{-1}$, corresponding to a mass of 1.04 $\times$ 10$^{8}$ M$_{\odot}$.
Its position and velocity approximate the end of the tidal tail in the east or a part of the Northwest Plume in the west, but neither is strictly continuous as shown in Fig.~\ref{fig:channel_map} and Fig.~\ref{fig:pv} (f).
The presence of the double-peaked spectral profiles at both joints, as well as its higher flux, appear to indicate that it is a separate feature.

\subsection{Probable gas of M51b}

\begin{figure*}
	\includegraphics[width=1.5\columnwidth]{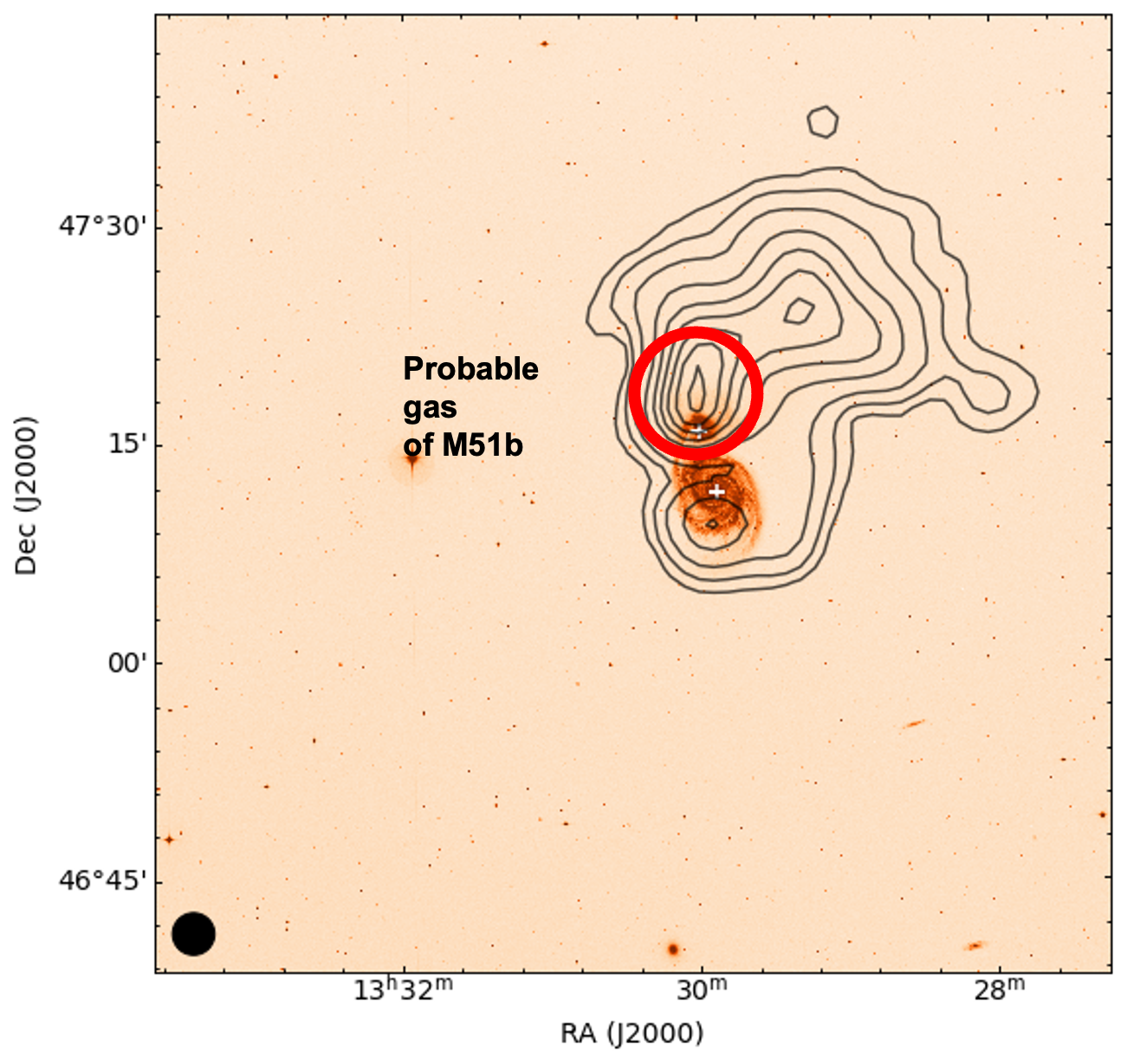}
    \caption{\ion{H}{i} column density contours integrated over 560-740 km s$^{-1}$ overlaid on the DSS B-band optical image.
    The contour levels are 2.55, 5.1, 10.2, 20.4, 30.6, 40.8, 51, 61.2, and 71.4 $\times$ 10$^{18}$ cm$^{-2}$.
    The red circle marks the location of the probable gas of M51b.
    The locations of M51a and M51b are marked by two white crosses, and the HPBW of FAST is indicated at the bottom-left corner.}
    \label{fig:560-740}
\end{figure*}

In addition to the features at the outskirts of M51 discussed above, there is a cloud of high velocity gas near the M51b location.
Although it had been observed before \citep{Rots1990, Walter2008}, it was too faint to attract attention.
We integrated the data from 560 to 740 km s$^{-1}$ and overlaid its contours on the Digitized Sky Survey (DSS) B-band optical image, as seen in Fig.~\ref{fig:560-740}.
The contour levels are 2.55, 5.1, 10.2, 20.4, 30.6, 40.8, 51, 61.2, and 71.4 $\times$ 10$^{18}$ cm$^{-2}$.
The gas marked with red circle in the picture is what we are interested in.
Its flux is about 7.5 Jy km s$^{-1}$, corresponding to a mass of 1.3 $\times$ 10$^{8}$ M$_{\odot}$.
We consider this gas to be the possible gas of M51b.
Firstly, its location is relatively close to the M51b optical centre.
The blue vertical line in Fig.~\ref{fig:pv} (e) represents the position of M51b's centre, which matches perfectly to the gas's edge. 
Second, its high velocity is substantially different from the disc velocity of M51a, indicating that it is not M51a's gas.
Third, the gap shown in Fig.~\ref{fig:channel_map} between it and the Northwest Plume implies that it is not part of the Northwest Plume.

\section{Discussion and Conclusions}
\label{sec:discussion}

Numerical simulation is an essential tool for studying the M51 system's interaction history. 
As further observations of the system were made, two simulation scenarios emerged: one with single-encounter \citep{Toomre1972, Hernquist1990, Durrell2003} and the other with multiple-encounter \citep{Howard1990, Salo2001, Theis2003, Dobbs2010}.
In this section we will compare the newly discovered \ion{H}{i} features with existing simulations and discuss their possible formation processes.

The Southwest Plume shows only as faint and scattered clouds in VLA observation due to their relatively low sensitivity and has no optical counterpart, rendering this feature absent from existing simulations, both for single- and multiple-encounter scenarios.
There is a low surface brightness galaxy IC4263 at the centre of plume.
However, its large red shift (z = 0.008950, \citet{Pature2002}) makes the Southwest Plume not belong to IC4263.
Such \ion{H}{i} clouds, which no optical counterparts, are often found in the vicinity of interacting systems, such as the Leo Ring \citep{Scheider1989}, HIPASS J0731–69 \citep{Ryder2001} and VIRGO\ion{H}{i}21 \citep{Davies2004}.
The Southwest Plume might be the result of tidal or ram-pressure stripping. 
Ram-pressure stripping occurs mostly in dense environments such as galaxy group or cluster \citep{Gunn1972, Boselli2022}. 
And the Southwest Plume's complex morphology and kinematics indicate that it is caused by tidal stripping rather than ram-pressure stripping.
Numerical simulations have demonstrated the possibility of massive \ion{H}{i} clouds without optical counterparts arising from tidal stripping scenarios and as high-density portion of \ion{H}{i} rings or arcs \citep{Bekki2005,Bekki2005b}.
In several simulations \citep{Salo2001, Durrell2003, Dobbs2010}, the Southwest Plume is in the orbit of M51b, which may be a region of high density in the tail or ring structure formed by M51b.
On the other hand, due to its correlation with M51a in position and velocity, it cannot be ruled out that it originated from M51a. 
In the latter scenario, its opposite direction to the Southeast Tail suggests a multiple-encounter scenario.

Because of the presence of optical counterpart and old stars, the Northwest Plume should be a result of the system's interaction.
Many simulations have comparable features, which are more similar in single-encounter scenarios than in multiple-encounter.
In \citet{Durrell2003} simulation, there is some structure in the northwest that is fairly close to our observation.
Since they observed planetary nebulae (PNe) of both velocity modes in this region, they suggest that the region is composed of the northwest tail of M51a and the west tail of M51b. 
The large velocity range span (200 km s$^{-1}$) of the Northwest Plume seems to be evidence for this view.
The majority of other simulations assume that this plume originates from M51a, a northwest tail corresponding to the Southeast Tail.
During the system interactions, the material close to M51b suffers the greatest dispersed tidal forces and is ejected as the tidal plume.
On the other hand, according to velocity continuity, it is also likely to come from M51b, as seen in Fig.~\ref{fig:pv} (e).
This scenario is similar to that of the NGC2782 system \citep{Smith1994, Onaka2018}, which contains a stellar tail to the east and a \ion{H}{i} tail to the opposite west. 
This phenomenon can be reproduced by a off-centre collision of two galaxies with a mass ratio of 1:4, with the massive galaxy forming the \ion{H}{i} tidal tail on the west side and the stellar tail on the east side coming from the small mass galaxy.
For the M51 system, the Northwest Plume is a remnant of M51b after a off-centre collision of the M51 system.
However, this scenario does not resemble any of the existing models for M51, as it is difficult to reproduce the morphological and dynamical properties of the Southeast Tail of M51a.

\begin{figure*}
	\includegraphics[width=2\columnwidth]{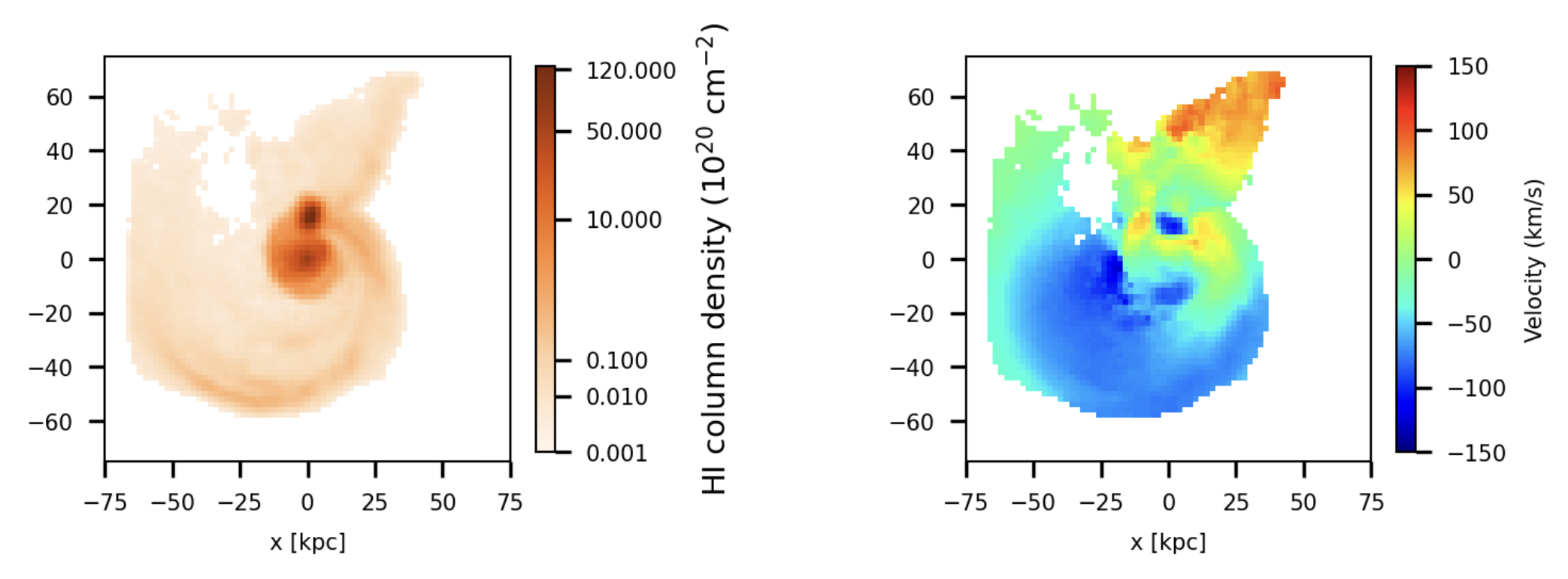}
    \caption{Gas properties of gas-rich galactic interaction $\sim$250 Myr after the pericentre passage, with a mass ratio of 3:1. 
    The left panel is a column density plot.
    The right is a velocity field plot.}
    \label{fig:simu}
\end{figure*}

In order to understand the origin of the Northwest Plume, we inspected the simulation of galaxy mergers carried out by \citet{Sauvaget2018}.
They have investigated a grid of representative simulations of gas-rich major mergers with mass ratio of 3:1. 
We found one simulation of polar orbit\footnote{
Specifically, it is referred to the model "POLAR-PRORET": a collision between two disc galaxies on a designed parabolic orbit with a pericentre of 16 kpc. 
"POLAR-PRORET" means both galaxies have their disc highly inclined (71 degree) to the orbital plane ("PROLAR") and the first galaxy (i.e., the large galaxy) has a PROgrade disc rotation comparing to the angular momentum of the orbit while the second galaxy has a RETrograde rotation. 
The total initial baryon mass of the large galaxy is $5.3\times 10^{10} $~M$_{\odot}$ (52\% in gas) and $1.76\times 10^{10} $~M$_{\odot}$ (72\% in gas) for the small galaxy. 
Both galaxies are embedded in a dark matter halo respectively, assuming a dark-to-baryon mass ratio of 4. } 
can well explain the observed dynamics in M51 system, as shown in Figure~\ref{fig:simu}, for example, the formation of the very extended \ion{H}{i} tidal tail of M51 as well as the global velocity field revealed by our FAST observation.
The wide and long tidal tail formed by massive galaxy and the high velocity tidal tail formed by small one on the opposite side are both similar to our observation.
Although the simulation does not fit all the observed properties of M51 system in details, it strongly suggests that the Northwest Plume is a tidal tail of M51b, which is analog to the further part of the tail of M51a. 
This feature, inspired from the simulation, implies the progenitors of M51a and M51b are all gas-rich and proceed a relative larger gas disc than their optical disc.

In addition, the Northeast and Northwest Clouds are also relatively distinctive tidal features. 
Although there is some spatial velocity overlap with the Northeast Plume, the lower velocity and higher flux make the Northeast Cloud appear to be a separate tidal feature with no optical counterpart. 
The Northeast Cloud is not present in the current simulation. 
It might be the tidal feature in the scenario of the system's multiple-encounter, or the gas expelled by other processes.
The Northwest Cloud is present as an optical counterpart and is represented in several simulations, corresponding to the northwest tidal tail of M51a \citep{Dobbs2010}. 
In addition to this, it is also recognized as a possible third part of the M51 system.

In summary, the high sensitivity of FAST allows us to recover large amount of diffuse neutral gas and identify several new extended tidal features outside M51. We classified the extended structure outside the M51 disc as the Southeast Tail, Southwest Plume, Northwest Plume, Northeast Cloud and Northwest Cloud.
The Southeast Tail, Northwest Plume and Northwest Cloud are represented in the simulations in both single- and multiple-encounter scenarios, but the Southwest Plume and Northeast Cloud are not. 
This might potentially be owing to the present simulations' limited resolution, so that it does not show the \ion{H}{i} gas at low column density. 
We also obtain a deep image of the neutral gas component that might be associated with M51b, with an estimated velocity range of 560 to 740 km s$^{-1}$ and a flux of 7.5 Jy km s$^{-1}$.
Because of the overlap in location with M51a's galactic disc, its dynamical properties will need to be examined in more details in conjunction with new simulations.
Both the newly discovered tidal features and the probable gas of M51b provide new insights into the M51 system and will contribute to the constraints on future simulation models. 
The generation of new models will contribute to the understanding of spiral structure, star formation and galaxy evolution in interacting systems.

\section*{Acknowledgements}

We thank the FAST staff for help with the FAST observations. 
We thank the reviewer and scientific editor for their constructive comments.
We acknowledge supports from the National Key R\&D Program of China (2018YFE0202900; 2017YFA0402600). 
We also thank Arnold Rots for sharing the VLA observation data on the web and Aaron Watkins for sharing the deep optical image.
YY thanks the National Natural Foundation of China (NSFC No. 12041302 and No. 11973042), also the support the International Research Program (IRP) Tianguan, which is an agreement between the CNRS, NAOC and the Yunnan University.

\section*{Data Availability}

The raw data used in the article will be published on the FAST website: \href{https://fast.bao.ac.cn}{https://fast.bao.ac.cn}.
The PID is N2021$\_$4.
Please contact the author (hyyu@nao.cas.cn) for processed data.



\bibliographystyle{mnras}
\bibliography{M51} 





\bsp	
\label{lastpage}
\end{document}